\documentclass{article}

\usepackage{arxiv}
\usepackage{amsmath}
\usepackage{amsfonts}
\usepackage{bm}
\usepackage[utf8]{inputenc} 
\usepackage[T1]{fontenc}    
\usepackage{hyperref}       
\usepackage{url}            
\usepackage{booktabs}       
\usepackage{amsfonts}       
\usepackage{nicefrac}       
\usepackage{microtype}      
\usepackage{lipsum}
\usepackage[font=small,labelfont=bf]{caption}
\usepackage{listings}
\usepackage{adjustbox}
\usepackage{cite}
\usepackage[ruled,vlined]{algorithm2e}

\usepackage{graphicx}
\usepackage{wrapfig}
\usepackage{xcolor}

\title{Scaffold-Induced Molecular Graph (SIMG): Effective Graph Sampling Methods for High-Throughput Computational Drug Discovery}

\usepackage{authblk}

\author{
  Austin Clyde, Ashka Shah, Max Zvyagin, Arvind Ramanathan, Rick Stevens \\
  Computing, Environment, and Life Sciences Division \\
  Argonne National Laboratory\\
  Lemont, IL 60439 \\
}

\begin{document}
\maketitle
\begin{abstract} 
Scaffold based drug discovery (SBDD) is a technique for drug discovery which pins chemical scaffolds as the framework of design. Scaffolds, or molecular frameworks, organize the design of compounds into local neighborhoods. We formalize scaffold based drug discovery into a network design. Utilizing docking data from SARS-CoV-2 virtual screening studies and JAK2 kinase assay data, we showcase how a scaffold based conception of chemical space is intuitive for design. Lastly, we highlight the utility of scaffold based networks for chemical space as a potential solution to the intractable enumeration problem of chemical space by working inductively on local neighborhoods.
\end{abstract}


\section{Introduction} 
The multi-step process of drug discovery is incredibly resource intensive. The time it takes to go from initial compound search to a clinically tested, effective product can range between 10 to 15 years\cite{Hughes_et_al:2010}. The estimated cost of research and development that leads to a viable drug in the market  today is about US\$2.5-3 billion \cite{saha2020pharmaceutical}. Considering the universe of about $10^{60}$ possible compounds to traverse for effective drugs, we seek to enable more efficient, higher throughput, and more meaningful frameworks for compound analysis \cite{Bohacek_et_al:2010}. The ongoing COVID-19 pandemic underscores the urgent need for effective \emph{sampling} strategies to sample large chemical spaces and identify small molecules that can target the SARS-CoV-2 proteome.    

A number of novel approaches have already been developed to accelerate the drug discovery pipeline. Advances in computational methods and hardware, coupled with investments in data mining for target identification and combinatorial chemistry over the past 20 years now allow us to mine extensive compound libraries and apply advanced modelling techniques for tasks like lead generation and optimization \cite{Yang_et_al:2011}. Alongside considerable advances in the high throughput screening (HTS) space, the magnitude of interesting leads we are able to generate prompts a growing body of work on novel virtual screening (VS) techniques \cite{Oprea_et_al:2004}. Due to the physical and financial constraints of automated HTS systems, new generation and VS methods are key to scaling research so that billions of new and interesting compounds can be considered and evaluated. 

This desire to scale up and deliver deeper insight into the potential efficacy of new molecules has prompted the integration of high performance computing and HTS methods in more recent years. For example, Lyu et al. evidence the promise of extensive screening of full compound libraries to discover their most biologically active compounds \cite{lyu2019ultra}.  They docked 99 million molecules and then clustered the resulting top-ranked 1 million molecules. After removing compounds similar to known inhibitors and reducing redundancy, they identify fifty one top-ranking molecules, 86\% of which are successfully synthesized, and 5 of which are measured to effectively inhibit for an overall hit rate of 11\%.  In some ways, the key to their success was  increasing the library size to 100 million, going beyond the standard chemotypes and assays. Other works have focused on scaling, as well as increasing the accuracy of st{the} computational screening \cite{saadi2020impeccable}.

With the rise of standardized synthetically feasible small molecule libraries, computational and medicinal chemists have noted the pre-existing small compound libraries used for HTS in past decades are biased towards certain properties, many of which were useful without access to state of the art computing and methodology \cite{jia2019anthropogenic}.  Accessing these new large-scale libraries is an exciting area with unbounded potential for new pharmaceuticals and therapies. Even with the slew of upcoming exascale machines, current computational tools are limited in their ability to screen billions of molecules. If a computational technique existed with the positive detection power to pull out a manageable subset of compounds from a library of 100 billion compounds, the scale of deployment and diversity of compounds explored will always be bounded by storage limits of enumerating a database and the detection power of the computational models. Current models aimed to accelerate ultra-high-throughput screening (uHTS) drug screening computations are only able to successfully filter datasets by one to two orders of magnitude \cite{clyde2020regression}. 

Current approaches rely on an exhaustive search of compound libraries with no inherent structure. Given computational techniques such as docking are not getting faster, GPU accelerated machine learning has taken precedence \cite{kaushik2020ai,schneider2020rethinking}. While this has certainly expanded the size of accessible chemical space by an order of magnitude, the technique remains bounded by GPU acceleration.  In order to reach the scale and diversity desired for uHTS screening, a new way of conceptualizing chemical space is required. In this paper, we propose a novel sampling technique that is not tied to computational power and provides unrestricted access to chemical space without enumeration. 


Sampling the chemical space defined by existing drug-like molecule databases and beyond remains a challenge. Some researchers have described the space in various ways in an attempt to make sampling more useful for screening \cite{reymond2012exploring, oprea2001chemography}. Schwartz et al. use the idea of a property space as a geometrical realization of the concept of the chemical space by positioning molecules in a multidimensional Euclidean space according to a selected set of properties/descriptors. They develop a similarity search algorithm based on SMILES fingerprints (SMIfp) \cite{schwartz2013smifp}. Hall et al. show that fragment based libraries have better coverage of the chemical space than larger molecule libraries by analyzing the topology of commercially available fragment libraries. They argue therefore that fragments serve as good probes of the chemical space \cite{hall2014efficient}.

Others have developed techniques for sampling outside existing databases. Virshup et al. use a stochastic search to create new structures using chemical mutations, then prune according to undesirable properties with the goal of allowing systematic exploration of uncharted chemical space \cite{virshup2013stochastic}. Several machine learning approaches have been developed for sampling the chemical space. Smith et al. use active learning to generate datasets for predicting molecular potentials for organic molecules. Compounds that have high variance of ensemble predictions are selected to initiate search of new molecules/conformations and can be used to sample high error regions of the chemical space automatically \cite{smith2018}. G\'{o}mez-Bombarelli et al. use an autoencoder to create a continuous latent representation of the molecule; the representation is then optimized for certain desired properties \cite{gomez2018automatic}.

In the pursuit of exploring the novel chemotypes in these libraries, the standard computational practice will not be sustainable beyond the current sizes of data available. With the current approach, every compound is typically run through inference property models followed by docking/physics based modeling. While GPU computing has made ML inference very fast, it still would not be able to screen anywhere near the imaginable small molecule space ($\sim10^{60}$). In the short term, high throughput docking/property prediction will be a useful technique for identifying active compounds in current libraries. Looking forward, a successful approach has to rethink the exhaustive search bottlenecking drug discovery. Facing the reality that naive exhaustive inference predictions or docking alone will not be able to touch most of chemical space, novel sampling methods must be developed to move beyond standard uHTS drug screening datasets. 

We posit that a simple, yet powerful graph-based abstraction of the molecular space can result in effective sampling methods to explore (and exploit) the chemical diversity to identify small molecules that effectively bind specifically to a protein target. Our approach draws inspiration from how medicinal chemists typically approach the problem of optimizing small molecules to bind to the protein's active site. This process relies on the fact that there are some known scaffolds (typically sub-graphs of known molecules) that form stable interactions with specific active site protein residues followed by  ``decorations'' to these molecular scaffolds that constitute certain functional groups. These transformations can be modeled as a graph, where nodes represent chemical compounds, and edges corresponding to the decorations (i.e., transformations). The addition of favorable functional groups can induce better binding, where as other groups can potentially induce clashes, reducing the ability for the molecule to bind in the active site. We can view this as a scaffold induced subgraph where some chemical scaffolds induce better interactions and others need not, leading to a set of discrete transformations that enable us to successively transform a scaffold `primitive' into a small molecule that ends up binding to the protein in a specific manner. This process effectively exploits graph representations of small molecules, where one can imagine hoping or walking among different functional groups or induced scaffolds.

Further, we explore the natural inclination to use a predictive model over scaffold graphs. We apply a label propagation algorithm to predict the docking score of a scaffold graph of the Mcule database in an effort to prove smoothness of docking scores over the graph and provide a method to further minimize the need for docking significant subsets of large databases \cite{mcule}. However, we find that exploiting inherent graph structure and scaffold relationships for sampling is more likely to provide better enrichment than label propagation. 

We present preliminary results in applying this approach conceptually and experimentally. First, we show using activity and docking data from \cite{clyde2021high} that embedding molecules into hypergraphs induces a natural clustering which corresponds well to the underlying data. Second, we show that docking 0.1\% of a chemical database using random walks based on scaffolds can identify the class of compounds of the highest scoring compound without needing to dock anywhere near the whole dataset. Lastly, we highlight the use of some standard predictive tests that attempt to use the scaffold relationship to smooth predictive errors. 

\section{Methods} 
We first outline the basic application of hypergraphs to molecular databases. We then outline two applications, first comparisons of graph topology to elucidate a new understanding of chemical space, and second, graphs for learning problems with low data. 

\subsection{Graphs}
We utilize a standard notation of graphs. A graph is a tuple $G=(V,E)$ where $V$ is a set of nodes and $E$ contains pairs of nodes. A directed graph means $E$ contains \textit{ordered} pairs, while an undirected graph's edge set contains unordered pairs (i.e. if $(e_1,e_2)$ is in $E$ then $(e_2,e_1)$ is in $E$).  A hypergraph is a generalization of a graph where edges contain more than two nodes ($E$ is a subset of the power set of $V$). 

\subsubsection{Graph Relations on Molecules}
In order to induce a graph structure on molecules, a relation is needed so given two (or more) molecules we can determine if there is a relation (edge) between the two. The generality of this relation is useful as different sciences have different theoretical priorities in their view of chemical space. For instance, we say molecule $X$ has fragment inclusion with molecule $Y$ if a fragment of molecule $X$ is contained in molecule $Y$. This type of relation is often used to perceive chemical space in fragment based drug discovery \cite{kirsch2019concepts}. In polymer studies, such a technique would not be fruitful as a relation on a monomeric unit captures more faithfully how a polymer scientist would view chemical space \cite{judzewitsch2020high}. 

In this paper, we adopt the notion of Bemis-Murcko scaffolds \cite{bemis1996properties}. The definition of a chemical scaffold emerged as a conceptualization of the skeleton or core in the late 1990s \cite{schneider1999scaffold, bemis1996properties}. It has proven to be an effective tool in finding ``isofunctional molecules" which can be used in the same way as an original starting molecule, while differing significantly in physical composition. This approach is referred to as scaffold-hopping, and is a well-established technique in lead generation and drug discovery \cite{sun2012classification, mauser2008recent}. Bemis and Murcko define a ring system as one or more rings connected via an edge, and a framework consists of all ring systems and linkers of a molecule. In the original paper, they did not include unsaturated hetero bonds on the main framework, however due to our interest in protein-ligand binding, our fragmentation includes unsaturated hetero bonds on main framework. Furthermore, on these frameworks, we can define a simple rule for producing smaller fragments to produce a network. Given a framework or scaffold, one can break the linker, and unfuse ring systems. The hierarchy of a scaffold is the number of ringers, and the hierarchy is additive (i.e. a scaffold in hierarchy three is necessarily composed of a scaffold from a two ring system and a one ring system combined either through sharing an edge on the ring systems or through a linker, or similarly three hierarchy one scaffolds). 

Expanding these relations to a collection of molecules yields a chemical scaffold graph. However, Beyond being used to perform targeted searches for isofunctional matches, the concept of scaffolds can be combined with graph and network theory in order to organize a broad chemical space. This is done by creating a scaffold graph or network using scaffold decomposition. Each molecule within a preexisting chemical space can be broken down into its corresponding scaffold tree, and the set of molecules can then be related to each other based on these common scaffold components. We utilize the open-source Python package available named \texttt{ScaffoldGraph}, which constructs a scaffold graph from a list of SMILES on which we chose to organize our chemical search space \cite{scott2020scaffoldgraph}. An RDKit based implementation is also available \cite{kruger2020rdscaffoldnetwork}. This scaffold graph is the hypergraph on which we learn properties, such as docking score, for each node. The chemical scaffold graph can be created with a simple algorithm. Given a set of compounds, for each compound one computes the Bemis-Murcko scaffold. Both the molecule and their corresponding scaffolds are added as nodes to the graph, and the edges from the scaffold to the molecule are added to the edge set. Next, for every scaffold that has more than one ring, break the scaffold into all possible pieces without breaking ring structures. These new smaller scaffolds are then added to the node set and edges are added connecting each of these smaller scaffolds to the scaffold from which it was derived (i.e. connect smaller structures to their super structures). This process is repeated until every scaffold has been broken down to single rings.

Once this scaffold graph has been generated, the docking scores from a docking program can be added as properties to the vertices of specific molecules. Now it is possible to assess the continuity of the scaffold graph with respect to the target property of binding affinity. This tests that the difference between two molecules' binding affinity is a function of the scaffold distance between them. This metric of continuity is evaluated using a semi-supervised classification method, where the actual binding scores are compared to those predicted by label propagation. If the docking score continuity is successfully established, then this allows for improvement in sub-sampling when performing screening of molecules within a chemical space. Instead of selecting a random set of molecules to screen for docking scores, or spending the resources to run the entire set, it is possible to then initially dock only the outer layer of the graph, and then use those results to drastically narrow down the search by propagating the scores inward.

\begin{figure}[!htbp]
  \centering
  \includegraphics[width=0.95\textwidth]{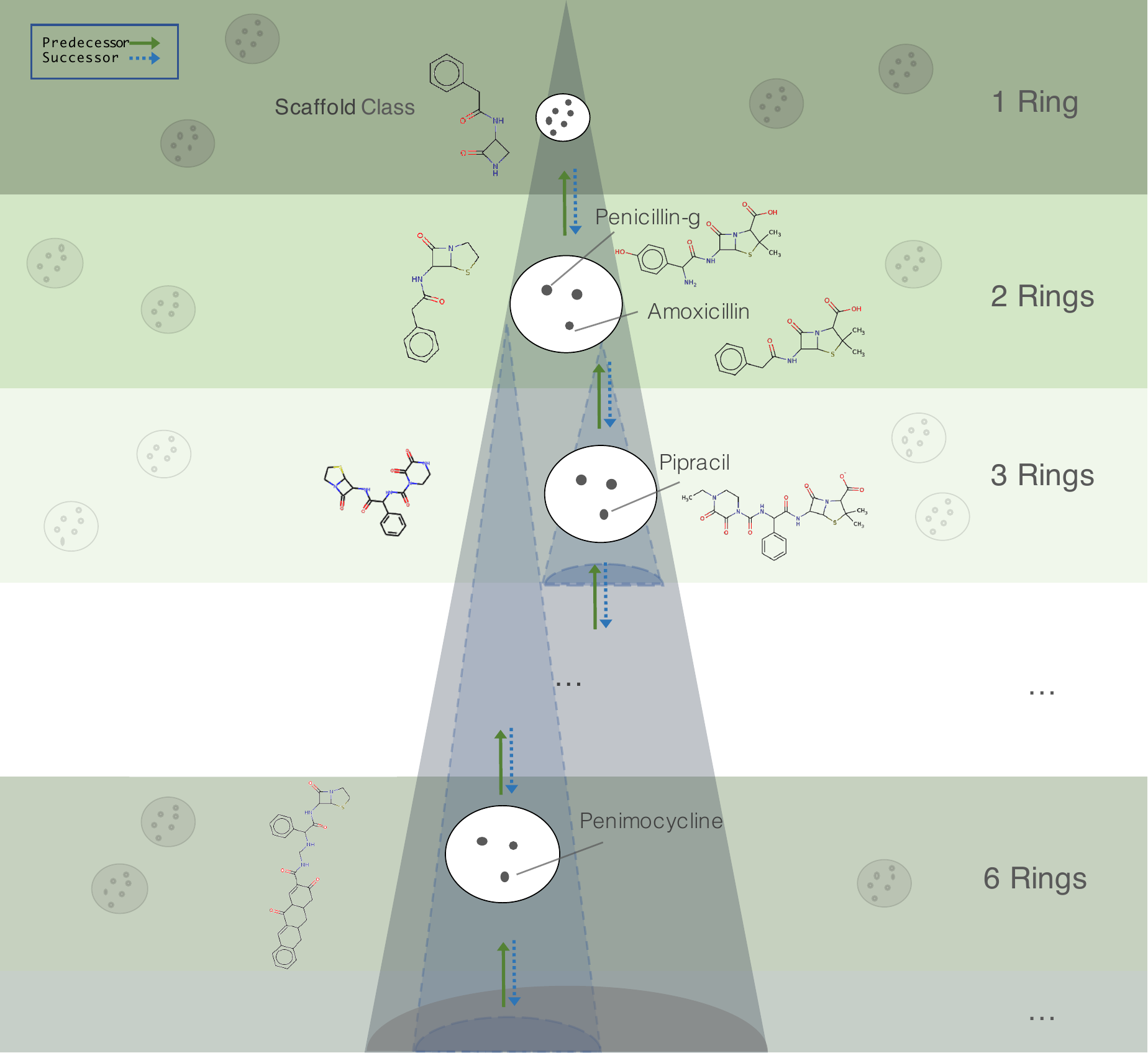}
  \caption{Scaffolds as classes over drug-like chemical space. Every
molecule (represented by dots or depiction inside circles) is
inside a single scaffold class. Scaffold classes are related through
common substructures, forming a hierarchy of classes. Penimocycline,
for example, is a distant successor of
Penicillin-g’s and Amoxicillin’s scaffold class, while Pipracil is a direct successor
of the same scaffold class. Lower cones, as depicted in the shaded regions, are defined as the sets of successor scaffold classes from any given class. Upper cones are the corresponding sets of predecessor scaffold classes. }
\label{fig:cones}
\end{figure}

\subsection{Fast Docking with Random Walks}
In order to sample from this graph to actively decide which molecules should be docked next, we need to carefully define how docking results will be applied to the sampling weights. There are various strategies for sampling nodes from a graph. The simplest strategy is random node sampling; however, this sampling strategy would not take advantage of our graph structure and would simply be the same as naive random screening molecules from a database. A more complicated strategy which incorporates the graph structure is random walks. 

Random walks are a stochastic process which defines a technique for taking a series of steps over a set. For example, a random walk over a graph (with nodes and edges) implies picking a starting node and randomly stepping in sequence from the starting node to one of its neighbors, then to one of the new node's neighbors, and so on and so forth. Random walks can be used as a sampling technique on graphs, where the samples are drawn by picking the nodes for which a random walker ``walks'' on.

The two essential pieces of information to define a random walk strategy is a starting node and a selection strategy to go from the current node to the next node. Given a graph $G=(V,E)$, where every $v\in V$ is a scaffold or molecule, we define a random walking strategy by setting the starting node to a random molecule and setting the walking strategy to be random half of the time and to select the neighbor with the best docking score the other half of time. In other words, half of the time the next molecule chosen will be a random neighbor of the current node, and the other times it will be a better scoring molecule. The use of both random walks and a particular objective function, in this case optimizing the docking score, is known as mixing. By utilizing mixing with the known docking scores we not only incorporate the graph structure we impose into the sample strategy, but we also incorporate structural information about the ligand into the search strategy. The goal is to avoid regions of chemical space which certainly would not contain hits (i.e. scaffolds which are merely the wrong shape for a particular binding pocket).

Let $\Phi_\theta:\mathcal{M}\rightarrow\mathbb{R}$ be a function mapping molecules to docking scores. Let $G_S=(V,E)$ be a molecular scaffold graph and let $H:V\rightarrow \mathcal{M}^<$ be a set function to produce the set of molecules that correspond to the scaffold from the data. The \textit{lower cone} of a scaffold $s\in V$ is a set $L_s\subset \mathcal{M}^<$ and defined as
\begin{equation}L_s=\{m : \text{t is a successor of s and } m\in H(t)\}.\end{equation}
Where a successor to a scaffold $s$ is defined as a scaffold in a higher hierarchy level (where higher levels contain more rings) with molecules that contain the structure of scaffold $s$. Informally, a lower cone is the set of successor classes from any given scaffold. Shown in Fig. \ref{fig:cones} is a visualization of how we conceptualize these cones in the chemical space of scaffold classes. The lower cone $L_s$ is a set essentially containing all molecules which share a common scaffold fragment, $s$. For each scaffold landed on during the random walk, we can sample either the molecules which strictly decorate that scaffold or we can consider all molecules which contain that scaffold (i.e. including larger scaffolds as well). In the result section, we show data utilizing the latter cone based conception given often times in drug discovery one fixes a particular scaffold component (such as an RNA-analog for proteases) and wants to find molecules which contain such structure.

\subsection{Learning on Hypergraphs}\label{sec:learning}
Given a graph $G=(V,E)$ let $F_V$ be node features so $V$ and $F_V$ have same length, which means each row of $F_V$ corresponds to a feature vector. Let $L\subseteq V$ be the subset of labeled nodes and $F_{VL}$ be the corresponding feature matrix. Let $U\subseteq V$ be the subset of unlabeled nodes and $F_{VU}$ be the corresponding feature matrix. We focus on regression tasks, where our goal is to make real-valued predictions of a given property for the unlabeled nodes $U$. Our algorithm first makes a baseline prediction with a neural network -- this step does not make any use of the graph $G$. Then we correct these predictions by propagating the residuals over the structure of $G$. In this step we assume graph smoothness and positive correlation of residuals along edges of the graph. This assumption of smoothness is by no means a guarantee for properties such as docking score over the structure of a scaffold graph. Instead, we hypothesize that the correctness of our final predictions for the nodes in $U$ using this method can confirm the validity of the assumption for a given property. The algorithm is outlined in Alg. \ref{alg:labelprop}: (i) Train a simple multi-layer perceptron (MLP) neural network $M$ on the subset $L$, using feature matrix $F_{VL}$ and labels $y_L$. Make predictions on the whole graph $\hat{y}=M(F_V).$
(ii) Estimate the residuals for the unlabeled data $r_U^{LP}$ using label propagation with the labeled residuals $r_L = y_L-\hat{y}_L$. Correct the predictions for the unlabeled nodes. This algorithm is taken from \cite{jia2020residual}, which demonstrates the effectiveness of label propagation with residuals on several regression tasks. 

\begin{algorithm}[H]
\SetAlgoLined
\SetKwData{Left}{left}\SetKwData{This}{this}\SetKwData{Up}{up}
\SetKwFunction{Union}{Union}\SetKwFunction{ConjugateGradient}{ConjugateGradient}
\SetKwInOut{Input}{input}\SetKwInOut{Output}{output}
\Input{Subset $L$ of labeled nodes, $\hat{y}$ predictions from model trained on $L$}
\Output{$\hat{y}_U^{LP}$, smoothed predictions from LP on residuals}
$r_L\leftarrow y_L - \hat{y}_L$\;
$\hat{r}_U^0\leftarrow$ $\bm{0}$ \tcp*{set initial guess for unlabeled residual's}
$\mathcal{L}\leftarrow \bm{I}- D^{-\frac{1}{2}}AD^{-\frac{1}{2}}$\tcp*{$A$ and $D$ are the adj. and deg. matrices} 
$\hat{r}_U^{LP}\leftarrow$  \ConjugateGradient($\mathcal{L}_{UU}$, $\mathcal{L}_{UL}r_L$, $\hat{r}_U^{0}$)\;
$\hat{y}_U^{LP}\leftarrow \hat{y}_U + \hat{r}_U^{LP}$
\caption{Label Propagation with Residuals}
\label{alg:labelprop}
\end{algorithm}

By decoupling feature-based learning and graph learning, we avoid the computational cost of training neural networks on large graphs while still benefiting from the inherent graph structure in our datasets using fast learning methods like label propagation. Our method for learning combines work from Huang et. al \cite{huang2020combining}, which uses a multi-step ``correct and smooth" after a base prediction from an MLP on classification tasks; and from Jia et. al\cite{jia2020residual}, which corrects using residuals on predictions from a graph neural network. Both show that using an initial base prediction with a model, and then post-processing using label propagation on the graph can boost performance and decrease training time.    

\section{Results} 
\begin{figure}[!htbp]
    \centering
    \includegraphics[width=5cm]{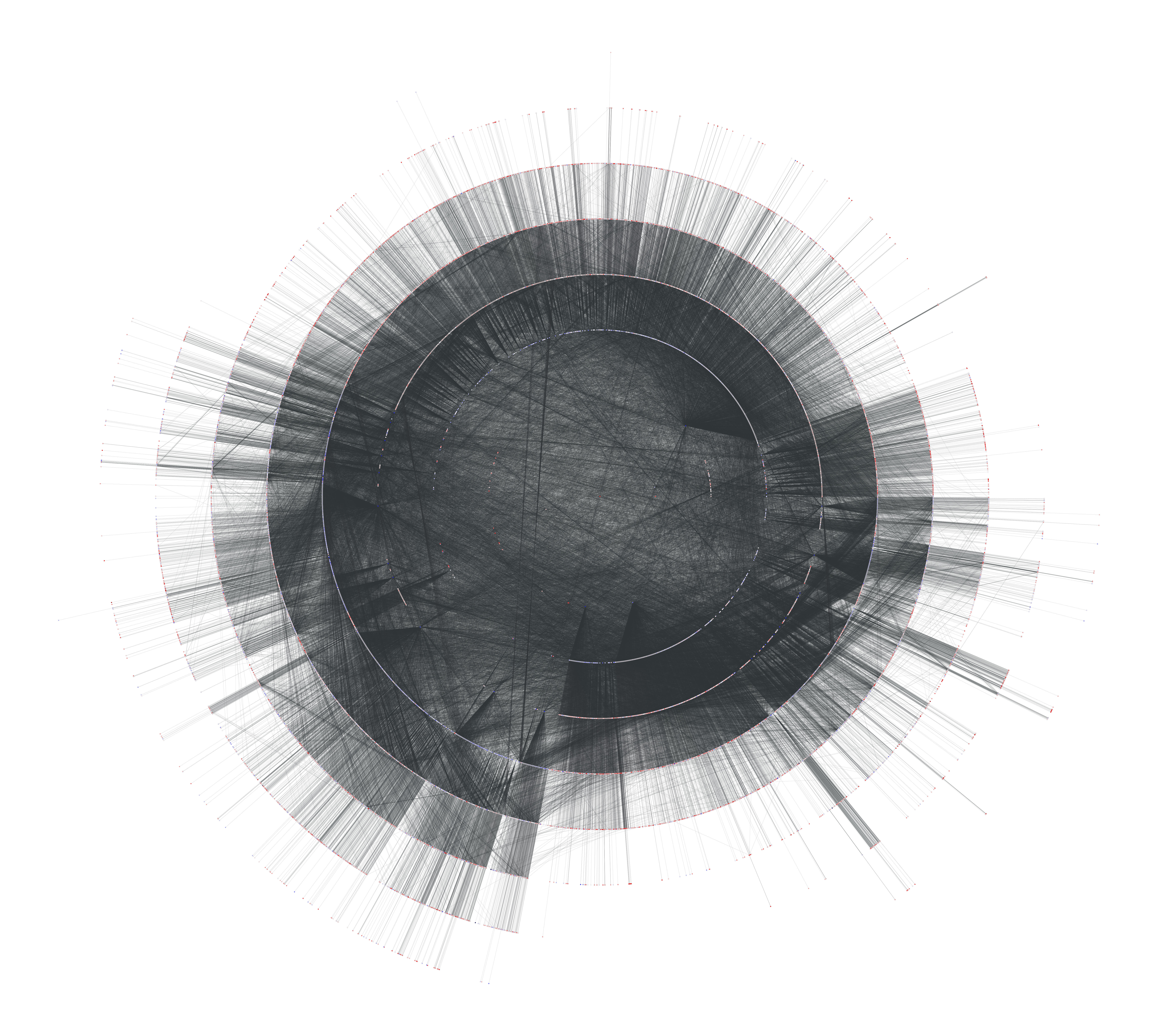}
    \includegraphics[width=5cm]{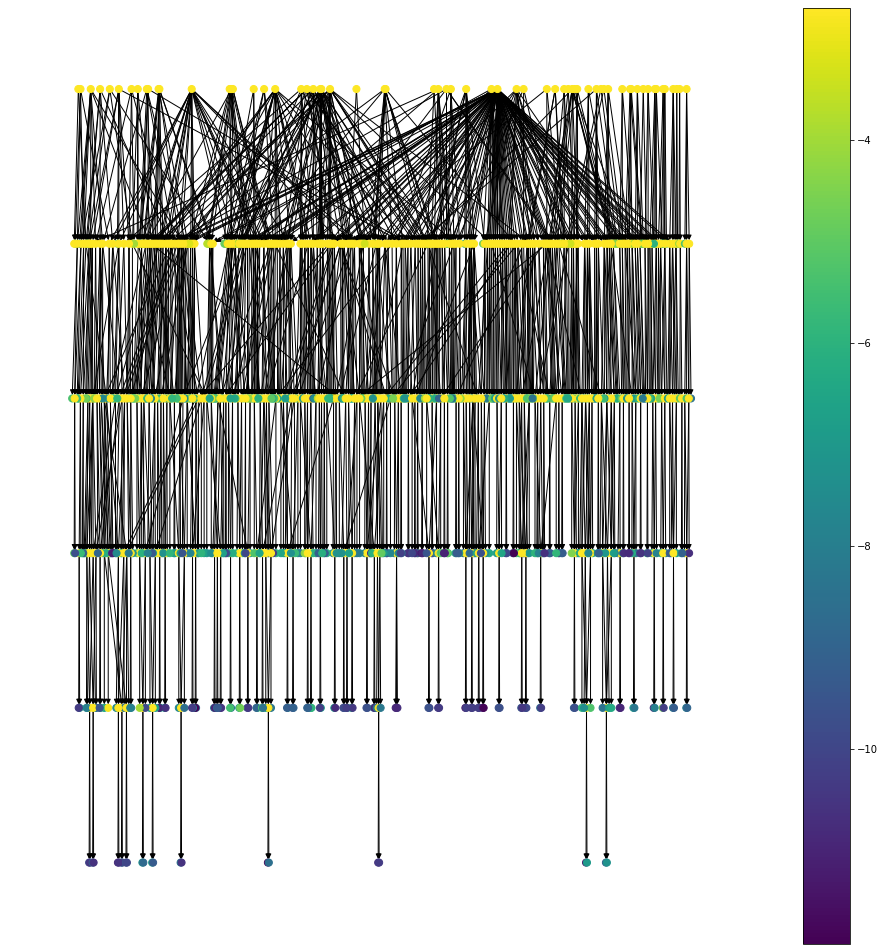}
    \caption{(left) {Radial graph layout of the Mcule molecular database. Pulling from these compound libraries typically results in a large connected component with a tree-like structure } (right) Subset of molecules structured in a tree graph used in COVID docking study \cite{babuji2020targeting}. The enumerated compounds from libraries typically appeared as terminal nodes (in purple shades), while the other nodes (scaffolds) connect those molecules together and provide a further in-depth view of the chemical space (yellow).}
    \label{fig:scaffoldgraph}
\end{figure}

\begin{figure}[!htbp]
    \centering
    \includegraphics[width=0.98\textwidth]{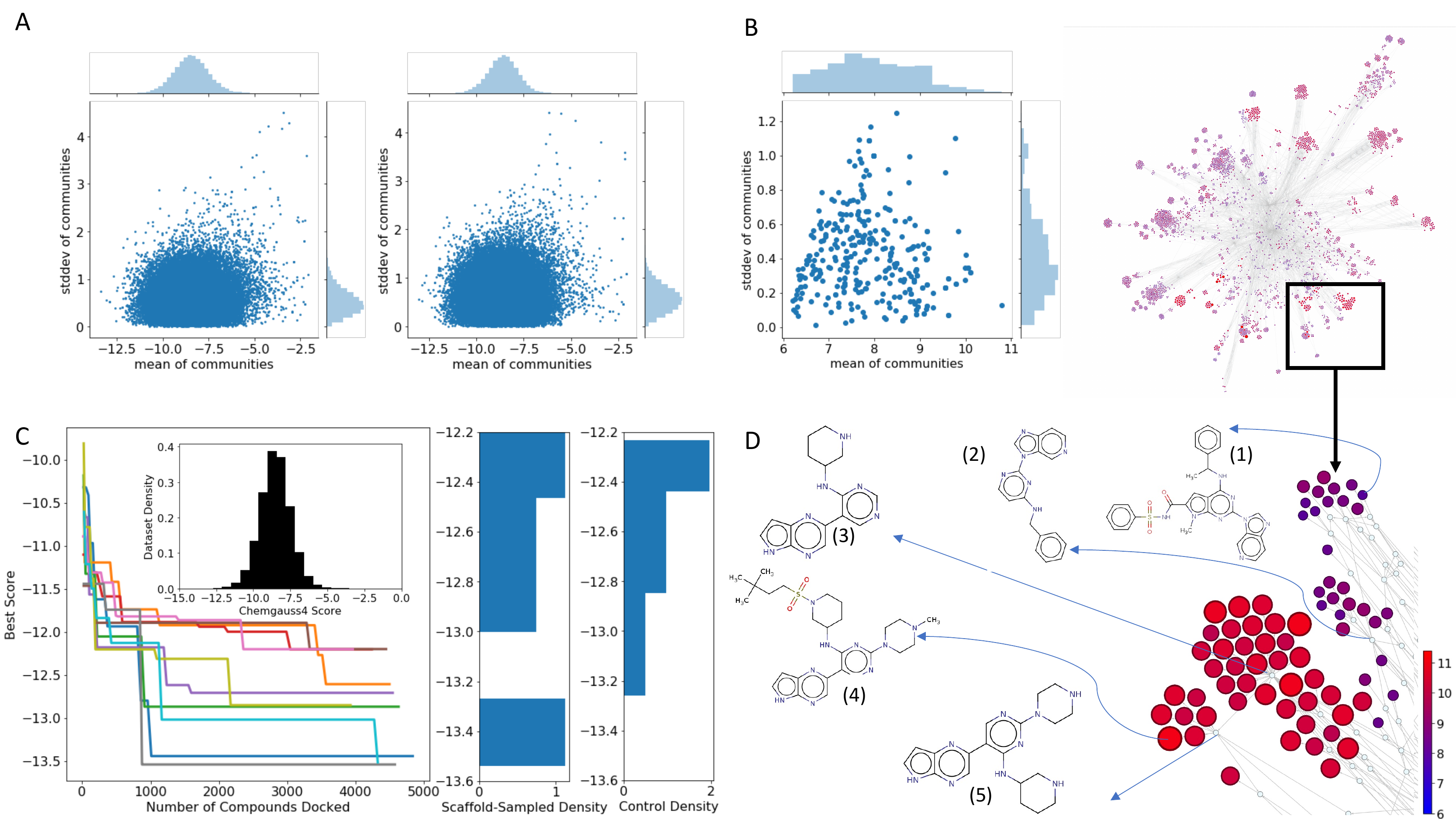}
    \caption{(A) Utilizing the scaffold graph based on the SARS-CoV-2 main protease (Mpro) dataset, communities contain all molecules which share a common substructure scaffolds. (left) this showcases the set of all communities based on three ring scaffolds, (right) the set of all communities based on four ring scaffolds. (B) The scaffold graph for JAK-2 kinase assay data from \cite{laufkotter2020kinase} (color bar in D). Red shaded nodes are near the top inhibitors in the data sample, and blue shaded nodes are near the bottom in the data sample. The grey shaded nodes are scaffold nodes which are not included in the data, but are used to organize the chemical space. (C) Highlights a docking simulation to show sampling based on the graph communities yields better performance than random sampling for capturing high performing compound classes with minimal docking (n=5000, 0.1\% of total data). (D) Zoomed in pane of B, showing similar compounds sharing a few common ring structures, but diverge in terms of assay performance (1 is $p=6.44$, 4 is $p=10.76.$).}
    \label{fig:mainresults}
\end{figure}

\subsection{Scaffold-space Partitions Virtual Screening Hits}
We generate a scaffold network for Mcule's in-stock compound library utilizing the ScaffoldGraph package \cite{mcule, scott2020scaffoldgraph}. The Mcule library consists of a diverse set of purchasable compounds within known stock amounts. Overall, the network has 6,369,219 nodes and 8,808,841 edges. The graph networks induced by the standard, drug screening libraries look as expected--high connectivity nodes include small one or two ring scaffolds with various pharmacophores, while terminal nodes are complex concatenations of the various predecessor graph nodes (Fig. \ref{fig:scaffoldgraph} left hand panel). 

Visualizing a large network with over 6 million nodes can be challenging -- hence, we chose a small subset of molecules for analysis (Fig. \ref{fig:scaffoldgraph} right hand panel). It is interesting to note that while many of the upper-level nodes in the graph (typically small functional groups such as benzene rings) have lower docking scores, successive transformations of these nodes towards the leaf edges have higher docking scores (favorable interactions). This means that traversing the graph from one radial end to the other can result in compounds that successively improve their overall ability to interact with the protein. Molecule nodes are only connected to their respective scaffold nodes, and scaffold nodes are connected based on substructure (i.e. scaffold node A is connected to scaffold node B if A is a substructure of B, and molecule node C is connected to node B if the scaffold of molecule node C is B). In a sense, this is simply a projection of a hypergraph to a regular graph. We assign molecule nodes a docking score based on a docking data set for SARS-CoV-2 main protease (Mpro) \cite{ClydeHTVS}.

Utilizing the notion of cones derived in the method section, Fig. \ref{fig:mainresults}A highlights the fact that molecules in the same cone are likely to have docking scores relatively close to each other. For the unstructured dataset, the mean docking score is 8.50 $\pm$ 1.05. The standard deviation of these cones is a proxy for the smoothness of the graph with respect to their docking scores. This is a good measure as it implies given another molecule with the same scaffold in that class, it is more likely that the molecule falls within the mean of the cone. The standard deviation of the cones is relatively small compared to the standard deviation of the overall data, and this difference is significant. This indicates that local scaffold neighborhoods are relatively uniform in their overall docking scores. We see that using scaffolds as a way of embedding molecules is natural and matches the intuition. 

We explore scaffold-space's relation with docking scores. Using a simple random walk which we outlined in the methods section, we show the strategy is more efficient in finding high scoring hits than random sampling. In Fig.\ref{fig:mainresults}C, the colored lines correspond to the best score seen so far by a random walker. The lowest gray line, for example, shows that after selecting 1000 nodes based on the random walk strategy, the lowest score found is -13.5. Conversely for the random strategy, the histogram in black in the zoomed-in pane shows that the lowest score found randomly was greater than -12.5. Therefore, random walking with mixing in docking scores is more efficient at sampling high scoring compounds in a small docking screen. We show that based on more activated scaffolds improves detection of the best docking scaffold classes over just randomly docking.

Scaffold-space is locally homogeneous with respect to experimental assay data. We illustrate JAK2 kinase data for approximately 3,000 compounds from \cite{laufkotter2020kinase} in Fig. \ref{fig:mainresults}B and D. The overall dataset reports p-values (higher indicates more likely inhibition) with a mean of 7.746 $\pm$ 1.03. We see that communities are much smoother than the entire dataset, with an average standard deviation half of the dataset as whole. In the chart of Fig. \ref{fig:mainresults}B, we see some high scoring communities (high $p$-values) have rather small standard deviations---this is illustrated in the graph depicted, where red and large nodes have larger $p$-values than blue and small nodes. We zoom in on a particular set of four communities in Fig. \ref{fig:mainresults}D, two of which are highly active based on the data and one which is not. We observe that (2), (3), and (5) are scaffolds, where less active molecule (1) has (2) as its scaffold, (3) is a super-scaffold of scaffold (5), and active molecule (4) has (5) as a scaffold. We observe that all molecules in the study which contain substructure (5) are active, while all molecules which have (2) has a scaffold are less active.
 
\subsubsection{Experimental setup and label propagation to scaffold nodes}
 We trained a DeepChem \cite{Ramsundar-et-al-2019} Multitask Regressor network, a feed-forward multi-layer perceptron (MLP), with a 1,000 node hidden layer and ReLU activation function. Our feature matrix $F_V$ consists of 2,048 bit vector circular fingerprints for each molecule in Mcule ($F_{VL}$) and each scaffold in the generated scaffold graph ($F_{VU}$). Docking scores for each molecule in Mcule are generated using OpenEye FRED docking tool. The model is trained on the labeled Mcule nodes and is run in inference mode for the unlabeled scaffold nodes. We performed 5-fold cross validation using 80\% of the Mcule set as the training set for the MLP model and as the labeled nodes for label propagation using the conjugate gradient method. Results are shown in Table \ref{tab:label-prop-results}. We see that in this context label propagation of the residuals does not improve upon initial model performance. There may be several reasons for this, including the fact that our labeled dataset consists of only the outer ``ring" of nodes in the scaffold graph (Fig. \ref{fig:scaffoldgraph}) and therefore is not uniformly sampled from the graph. Instead, we note that results in the previous section provide empirical evidence of the smoothness of docking score over communities of scaffolds. Further, we have shown that docking based on active cones can provide the enrichment needed to more effectively sample from databases.   
 
 \begin{table}
 \caption{5-fold cross validation performance of the two step algorithm on test sets of labeled Mcule nodes and docking scores for 3CLPro}\label{tab:label-prop-results}
 \begin{tabular}{ |p{3cm}||p{2cm}|p{2cm}|  }

 \hline
Algorithm step& $r^2$ &MAE\\
 \hline
 \hline
 MLP Model & 0.58    & 0.55\\
 \hline
Label propagation&   0.57  & 0.58\\
 \hline
\end{tabular}
 \end{table}

\subsection{Application to JAK2 Kinase Inhibitor Discovery}
\begin{figure*}[!htbp]
    \centering
  \includegraphics[width=0.8\textwidth]{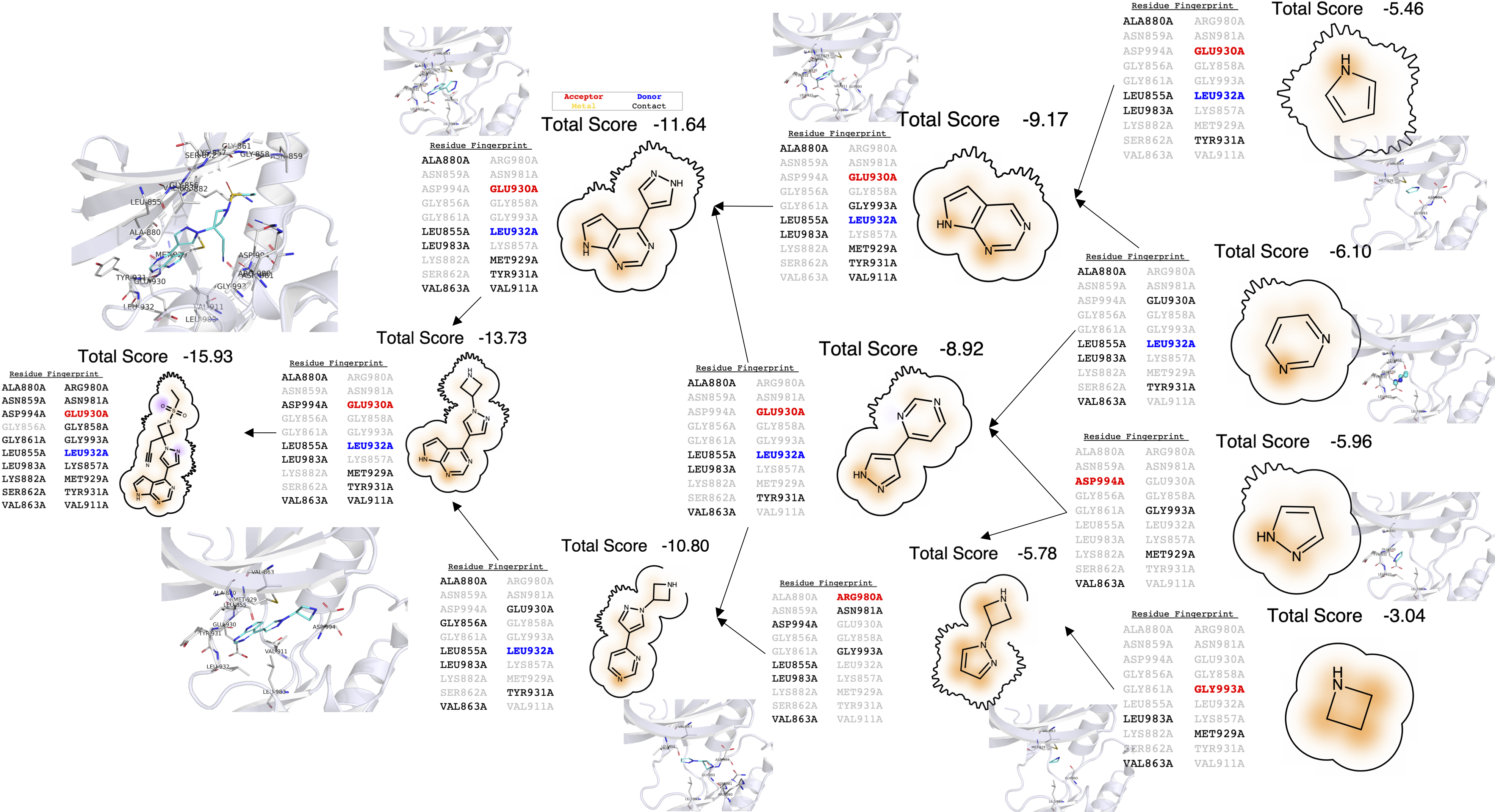}
  \caption{Baricitinib,a JAK2 inhibitor, (left) is decomposed by scaffolds into a graph (nodes being the molecules, edges showing how each decomposes to rudimentary single ring building blocks)  \cite{mcinnes2019comparison}. Each scaffold was docked as an independent molecule to JAK2 kinase (3KRR) using the FRED pipeline from OpenEye \cite{baffert2010potent, oechem2012openeye}. The residue fingerprints along with poses were also included. One can see how the scaffold based approach, on a microscopic level, shows promise as a meaningful conceptualization of chemical space---the merging of scaffolds with known properties leads to a compound with similar residue contacts.}\label{fig:jak2kinase}
\end{figure*}

This conceptualization of the chemical space for drug discovery has many applications for drug design. Fig. \ref{fig:jak2kinase} illustrates hits utility as a sampling method. For each descendent and common scaffold, interactions with the protein remain relatively stable. On the contrary, if descending down a certain path in the graph does not have the desired interactions, it is unlikely continuing down the path will yield higher activity. We are currently working on quantifying this graph smoothness property of chemical space in terms of commute and hit times \cite{chennubhotla2007signal}. This will provide bounds on how well the graph sampling notion outlined in this paper will accelerate giga-docking studies. Separately, by connecting chemical space topologically by scaffold, we are able to perform design as implied by network of residue contacts of docked scaffolds in Fig. \ref{fig:jak2kinase}.

\section{Discussion} 

We outline first the need for a new computational paradigm for cheminformatics workflows. We show how the introduced network based on scaffolds solves the enumeration problem cheminformatics faces. Lastly, we outline the future direction of utilizing scaffolds and networks in drug design.

\subsection{Computational Structures for Cheminformatics}
Current computational approaches to virtual drug screening and design are generally unstructured \cite{ruiz2017updated}. Computational scientists treat molecular space as a database listing. As a result of this simple but useful treatment of chemical space, interaction  between the AI/ML community and medicinal chemists has taken off \cite{elton2019deep}. This has been helped as well by the AI community's focus on keeping deep learning methods as hands off as possible \cite{zhang2018end}. 
But treating chemical space as tabular listings of structures will ultimately fail. The sheer size of chemical spaces makes listing it intractable, let alone performing any type of inference no matter how simple. The database approach to computing on chemicals will not last long without radical changes in computing architectures (inference over $10^{68}$ is completely intractable). To solve this fundamental problem with the current methodical theory, the community needs to move towards conceiving computational structures for chemical space which do not require listing and assume relationships between molecules.

Treating chemical space as a network solves the enumeration problem. By connecting molecule space and assuming some properties will be locally specific, nearly all molecules are only 10 hops away from each other. This paper proposes the start of this navigation paradigm, and there is certainly room to improve as we include the other aspects of a molecules' design as stated in \cite{bemis1996properties} (such as sidechains/electrostatics). Furthermore, the network allows chemical space to be conceived as an inductive database \cite{meo2005inductive}. An inductive network for chemical space would not require defining the graph at one time in memory, rather it would start from a particular set of molecules, and then utilize operations to grow and shrink the molecule space \cite{clyde2021scaffold, li2019deepscaffold, lim2020scaffold}. Furthermore, the global graph representation of chemical space is unique to other sampling methods and molecular generation methods as SIMG focuses on utilizing known compound scaffolds and designs, rather than completely novel chemotypes, and connecting those designs to the data through a graph of molecules. Molecular generation methods generate molecular designs often with novel or synthetically unfeasible chemotypes and scaffolds structures \cite{gao2020synthesizability}. Molecular generation techniques can be connected to the SIMG approach, by placing those novel compounds into a well ordered structured global graph. The SIMG technique does not preclude the use of novel structures, it provides a more ordered approach to integrating those chemical structures into the graph representation of chemical space. 

\subsection{Scaffold-space is smooth with respect to docking and protein-ligand experimental assays}
We found docking scores to be locally smooth around scaffolds. Docking scoring functions include shape complementary as a large component of the score. Given that the scaffold of a molecule mostly defines the shape, it makes sense docking scores vary smoothly with respect to scaffolds. In Fig. \ref{fig:mainresults}D, we see three scaffolds (2), (3), and (4). The coloring is $p$-values from a binding assay with JAK2. All three scaffolds share a common pyrimidine ring. Scaffold (2) has a \textit{3-pyrimidin-2-ylimidazo[4,5-c]pyridine} three ring system, which is different than the three ring system of (4) and (3). Further, we observe this difference alongside different local assay results. Molecules which have scaffolds (3) and (4) are active, while molecules which contain scaffold (2), such as molecule (1), are more likely inactive or less potent. This matches another study which focused on molecular generation constraining the scaffold class \cite{li2019deepscaffold}. They find ``the predicted activity against DRD2 for generated samples is significantly higher compared to random sampled molecules from ChEMBL, showing the model is good at utilizing privileged structures to generate bioactive molecules." This corroborates our Mpro docking results, which show that sampling the best seen, or ``privileged," scaffolds generates samples with higher activity as judged by docking score. As an observation, we also see that JAK2 kinase assay data forms local communities around scaffolds that are locally smooth. Other work has shown for yet another class of compounds that ``the binding mode of the pyrazole carboxylic ester scaffold [to phosphodiesterase] does not change when substituents are added" \cite{jhoti2007structure, card2005family}. This scaffold-local behavior is well studied, and is the essence of scaffold based drug discovery (SBDD) or scaffold hopping \cite{rabal2015novel,dimova2017computational,lai2020privileged}. 

While scaffold hopping alone has been successful, very interesting work has been done with altering the core of a molecule directly to hop from low to high activity regions. One such paper \cite{gjorgjieva2016discovery} switches the central core only of two molecular series, maintaining the electrostatics from sidechains, to obtain nM DNA gyrase B inhibitors. 
Ideally, our notion of chemical space would also include a specification for electrostatics to fully specify a molecule, rather than scaffolds alone. We believe this would more richly specify docking scores as it would account for variance seen by molecules with varying sidechains in the same scaffold class with different docking scores.  

Scaffold graphs can have immediate impact on how medicinal chemists utilize computing systems for their work. The presentation of scaffold-space matches current practices in medicinal chemistry for molecular series design \cite{lai2020privileged}. Future work will aim to extend this concept to complete system for SBDD. For example, users will be able to provide molecular data, and ask for automatic generation of molecular series for experiments based on sampling likely scaffolds. Scaffold constrained generation is not novel, and has been successful in various campaigns such as scaffold based generative models \cite{li2019deepscaffold, arus2020smiles}.

\section{Conclusions}
Our results suggest that a graph-based representation of chemical space libraries offers an automated and tractable approach to scaffold-based drug discovery. Our current results were illustrated on docking data for SARS-CoV main protease, and assay data for JAK2 kinase. For Mpro, we used docking data available to us through large-scale VS approaches to organize the chemical space and identify regions of the chemical space that may potentially lead to new molecules that can inhibit this protein.  These approaches suggest the complementary role that graph-based approaches combined with knowledge about targets can be used as a means to capture complex biological (protein-target specific) information and ultimately define methods to model protein-ligand design. 

From a computational perspective, we present a technique which solves the enumeration problem of chemical space. Utilizing a network as an inductive dataset allows exploration of all of chemical space, and only a local neighborhood needs to be stored and queried at a given moment of the design process. We believe this is a step forward towards automated drug design which is integrated with medicinal chemists' workflows and knowledge. We aim to continue pursuing a knowledge based discovery system for scaffold based drug design.

\subsection*{Funding}
 This research was supported by the Exascale Computing Project (17-SC-20-SC), a collaborative effort of the U.S. Department of Energy Office of Science and the National Nuclear Security Administration.

\bibliographystyle{unsrt}  
\bibliography{bmcarticle}

\end{document}